\documentstyle[epsfig]{elsart}
\begin{document}
\begin{frontmatter}
\title{Parity Effect in Mesoscopic and Nanoscopic Superconducting Particles}
\author{I.O. Kulik, H. Boyaci, and Z. Gedik}
\address{Department of Physics, Bilkent University\\
Bilkent 06533, Ankara, Turkey}
\begin{abstract}
Superconductivity in small metallic specimens is studied
with regard to the size dependence of the {\em parity gap}
($\Delta_P$), a parameter distinguishing between the energy
of even and odd number of electrons in the granule.
$\Delta_P$ is shown to be an {\em increasing} function of
level spacing $\delta$. The energy gap of superconductor
$\Delta$, on the other hand, {\em decreases} with
increasing $\delta$ and vanishes at $\delta=\delta_c$
which is of the order of $\Delta$. However, nonzero value
of $\Delta_P$ persists above $\delta_c$ in a {\em gapless}
superconducting-insulating state. Level degeneracy in small
specimens having perfect geometry changes the size
dependence of the parity gap, the Josephson effect, and
flux quantization. Parity gap is evaluated using an
interpolation procedure between the continuum limit
($\delta\ll\Delta$),
the moderate mesoscopic regime ($\delta\sim\Delta$),
and the nanoscopic scale
($\delta\gg\Delta$), for which an exact solution to the
pairing problem is provided with the numeric diagonalization
of system Hamiltonian in a small metallic cluster.
\end{abstract}
\end{frontmatter}
%
pacs[74.20.Fg,74.80.Bj,74.50.+r]

\section{Introduction}
\begin{figure}
\hspace{4cm} \psfig{file=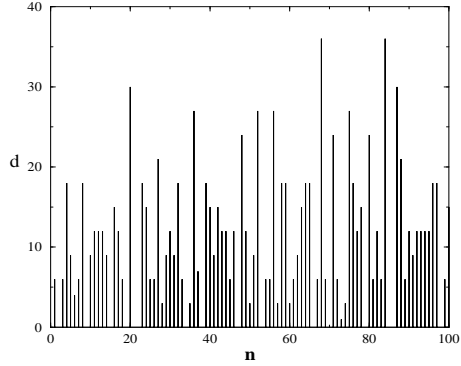,height=5cm,width=6cm}
\caption{Degeneracy of energy levels for parabolic energy vs
momentum dependence of total energy
$E_n=Cn^2=C(n_1^2+n_2^2+n_3^2)$.  \label{deg}}
\end{figure}

Superconductivity in metals is related to condensation of
electrons into pairs which are extended objects with a size
of order $\xi\sim\hbar v_F/\Delta_0$.
A fundamental question \cite{anderson} is
whether pairing can survive in small specimens with size $d$
less than $\xi$. In energy scale this corresponds to condition
that $\delta_1>\Delta_0$ where $\delta_1=\hbar v_F/d$ is typical
scale of dimensional energy quantization. The energy levels can
remain highly degenerate in a perfect sample (see Fig. \ref{deg})
but in a disordered sample, level spacing $\delta_2\sim \varepsilon_F/N_s$,
where $N_s$ is the total number of sites (which is of the order of
total number of electrons, $N$), is much smaller than $\delta_1$ and
can put a limit to superconducting pairing at
$\delta_2\sim\Delta_0$.
In recent experiments\cite{black,ralph},
signature of superconducting pairing was traced in ultrasmall
specimens down to sizes of 10 nm. The superconductivity
reveals itself through the number-parity effect, the dependence
of superconducting energy on whether the number of electrons in
sample is even or odd. Matveev and Larkin \cite{matveev} introduced
a {\em parity gap} $\Delta_P=\frac{1}{2}(E_{2n+1}+E_{2n-1})-E_{2n}$
where $E_N$ is superconducting energy vs number of electrons $N$,
to characterize that the paired state is preferable over the
state with odd number of electrons. In a number of papers, in
particular in the paper of Braun and  Von Delft \cite{braun}, $E_N$
was calculated assuming that in an odd-number sample, electron state
near the Fermi energy is blocked from pairing, similar to the way
in which finite-number systems are treated in nuclear physics
\cite{ring}. In present paper, we investigate the suppression of
superconductivity in a finite size system by using an approximation
similar to that of \cite{braun}, and also
in the opposite limit, by starting from atomic-scale
system (small atomic cluster) for which an exact solution of pairing
problem is provided, and trying to match small- and large-size
approximations through an interpolation procedure.

\section{Large-scale approximation}
In the Bardeen-Cooper-Schrieffer (BCS) approximation \cite{kulik1}
a variational density matrix
$\rho_{\alpha\beta}=\frac{1}{Z}\phi_\alpha^*
\exp(-(H-\mu N)/kT)\phi_\beta$
is exploited to obtain the ground state energy with trial wave functions
\begin{equation}
\phi_\alpha^{(1)}=\prod_{{\bf k}}(u_{\bf k}+v_{\bf k}
a_{{\bf k}\uparrow}^\dagger a_{-{\bf k}\downarrow}^\dagger)
|0\rangle
{\mbox{  and }}
\phi_\alpha^{(2)}=\prod_{{\bf k}\neq{\bf k}_F}(u_{\bf k}+v_{\bf k}
a_{{\bf k}\uparrow}^\dagger a_{-{\bf k}\downarrow}^\dagger)
a_{k_F}^\dagger|0\rangle
\label{bcs1}
\end{equation}
for even and odd number of electrons, respectively.
In the BCS pairing scheme (case 1), assuming equally
spaced levels with distance
$\delta$, we receive $\Delta_{2m}=\Delta_{2m+1}=\Delta_0$
whereas in the blocked pairing state (case 2) $\Delta_{2m}=\Delta_0$ and
$\Delta_{2m+1}=\Delta_0-\delta/2$.
The energy of superconductor at $T=0$ is calculated as
$\langle H-\mu N \rangle=\sum_n(\xi_n-\sqrt{\Delta^2+\xi_n^2})$ resulting in
a parity gap value $\Delta_P^{(1)}=\frac{\delta}{4}$
(for $\delta\ll\Delta_0$) in the first case, and
$\Delta_P^{(2)}\simeq\Delta_0\ln\frac{\omega_D\delta}{\Delta_0^2}+
\frac{1}{2}\delta$ in the second.
Energy of the blocked state is larger than that of
the BCS state and therefore there is no reason to believe that the former
is a better approximation for pairing. There is no direct relation of the
above calculation to the question that there is a large uncertainty in the
number of particles, and assertion that number of particles in
$\phi_\alpha^{(2)}$ is odd is merely an illusion.

\section{Josephson effect in mesoscopic metal particles}
Consider two identical small superconducting grains interacting
through the Hamiltonian $H_T=\sum_{nm}T_{nm}a_n^\dagger b_m+H.c.$
where $a_n^\dagger(a_n)$ creates (annihilates) an electron in the
first grain in state $n$. Similarly $b_n^\dagger$ and $b_n$ are
operators for the second grain. By introducing the phases $\phi_1$
and $\phi_2$ to the Bogoliubov $u, v$- coefficients in Eq.(1), we calculate
coupling energy due to $H_T$ as a function of $\phi_1$ and
$\phi_2$ in a standard way \cite{kulik1} $\Delta E=\Delta
E_S'-\Delta E_S\cos(\phi_1-\phi_2)$ where
\begin{equation}
\Delta E_S=\langle|T|^2\rangle\Delta^2\sum_{|\xi_n|,|\xi_m|<\omega_D}
\frac{1}{(\varepsilon_n+\varepsilon_m)\varepsilon_n\varepsilon_m}.
\end{equation}
\begin{figure}
\hspace{4cm}
\psfig{file=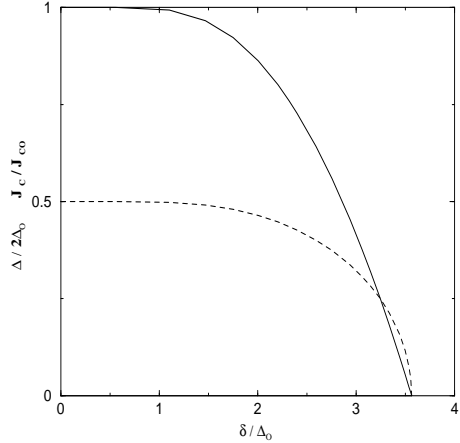,height=6cm,width=6cm} \caption{Josephson
critical current vs energy level spacing $\delta$ (solid line) and
gap parameter $\Delta$ vs $\delta$ (dashed line). The latter has
been scaled by a factor of $1/2$ for convenience. \label{jos}}
\end{figure}
The stationary Josephson current at zero temperature is
$J_c\sin(\phi_1-\phi_2)$ where $J_c=2e\Delta E_S/\hbar$. Figure \ref{jos}
shows dependence of Josephson critical current and energy gap
as a function of gap suppression parameter $\delta/\Delta_0$.
The curves show that superconductivity is suppressed at level spacing
$\delta_c=3.56\Delta_0$. This however is a conclusion within the mean field
approximation. It does not exclude the possibility that pairing may survive
for $\delta>\delta_c$ as a {\em gapless} superconducting state or as a state
of {\em superconducting insulator} with an effective insulator gap  $\delta$
larger than the superconducting pseudogap.
\begin{figure}
\hspace{2cm} \psfig{file=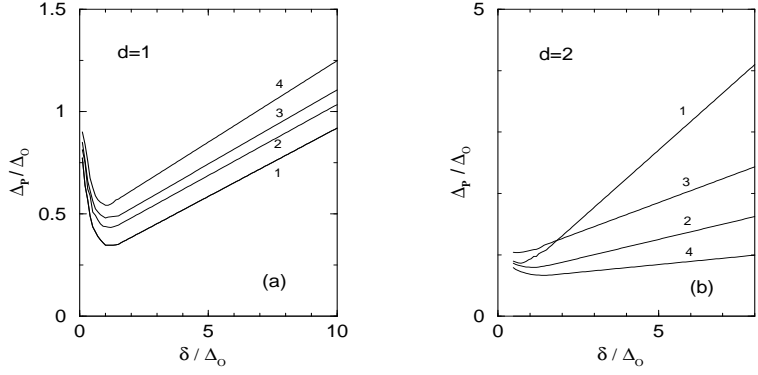,height=5cm,width=10cm}
\caption{Parity gap vs level spacing for
nondegenerate case (a) and for double degenerate levels (b). In
(a), numbers 1, 2, 3, 4 label dependences received with a number of
levels in the interaction shell $n_D=$30, 60, 120, and 360. In (b),
numbers 1, 2, 3, 4 correspond to index $i$ in various pairing gaps
defined according to
$\Delta_P^{(i)}=\frac{1}{2}(E_{4m+i}+E_{4m+i-2})-E_{4m+i-1}$. \label{boy}}
\end{figure}

\section{Richardson-Sherman approximation}
A finite size solution to the pairing problem was suggested by Richardson
and Sherman \cite{richardson} (and recently revived by von Delft \cite{delft})
assuming that states with energy within the interaction shell
$-\omega_D<\xi_n<\omega_D$ can be approximated as hard-core bosons whereas
states away from this shell are treated exactly as non-interacting particles.
The interaction Hamiltonian for hard core bosons is
\begin{equation}
H=2\delta\sum_{n\alpha}nc_{n\alpha}^\dagger c_{n\alpha}
-\frac{V\delta}{d}\sum_{|\xi_n|,|\xi_m|<\omega_D,\alpha\beta}
c_{n\alpha}^\dagger c_{m\beta}
\label{bos}
\end{equation}
where boson operators are
$c_{n\alpha}=a_{n\bar{\alpha}\downarrow}a_{n\alpha\uparrow}$, $d$ is level
degeneracy, and $\alpha$ labels states within the degenerate energy level
$\xi_n=n\delta$. Hamiltonian \ref{bos} is solved \cite{boyaci} by iteration
scheme proposed by Richardson and Sherman \cite{richardson}. Dependence of
$\Delta_P$ on $\delta/\Delta_0$ is shown in Fig. \ref{boy}.
In nondegenerate case ($d=1$) $\Delta_P(\delta)$ shows a behavior similar
to previous results with smaller number of levels\cite{mastellone,berger}.
In case of degenerate energy levels, pairing gap is multi-valued depending
upon which three successive states are used in calculation of $\Delta_P$.
The {\em true pairing gap} is possibly the largest one in Fig. \ref{boy}(b),
and others represent a new kind of mesoscopic effect which may or may not be
related to superconductivity. Small $\delta$ of $\Delta_P(\delta)$ does not
match with any of pairing gaps calculated in Section 3, which means possibly
that neither of the approximations (original BCS or BCS with one level
blocked) are good for small systems.
\begin{figure}
\hspace{2cm}
\psfig{file=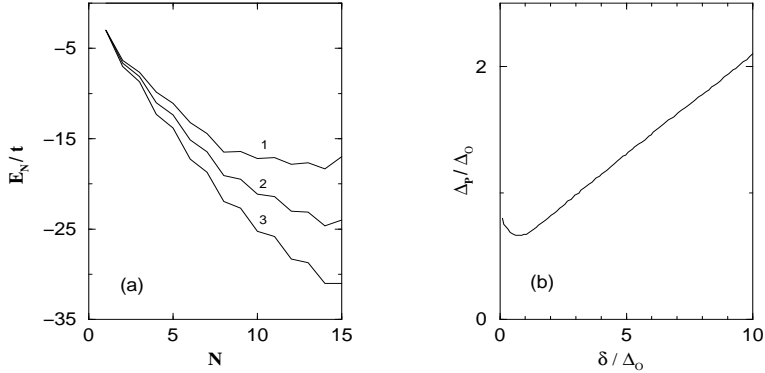,height=5cm,width=10cm} \caption{(a)Energy vs
number of electrons in cubic cluster with attractive interaction
at sites. (1) $U/t=2$, (2) $U/t=3$, (3) $U/t=4$. ~(b)Parity gap vs
level spacing in cubic cluster. \label{cls}}
\end{figure}

\section{Parity effect in atomic clusters}
We calculate the exact eigenstates of the negative-$U$ Hubbard Hamiltonian
\begin{equation}
H=-t\sum_{<i,j>,\sigma}a_{i\sigma}^\dagger a_{j\sigma}
-U\sum_{i=1}^N
a_{i\uparrow}^\dagger
a_{i\downarrow}^\dagger
a_{i\downarrow}
a_{i\uparrow}
+\sum_{i\sigma}V_ia_{i\sigma}^\dagger a_{i\sigma}
\label{hub}
\end{equation}
(with $U>0$) in small system of $N$ sites. $V_i$ is random potential at site
$i$. The fact that Hamiltonian \ref{hub} commutes with the total number of
electrons and also with the total spin leads to lowering of dimensionality
$2^{2N}\times2^{2N}$ of the matrix. Interaction is non-retarded (unlike
BCS Hamiltonian) and momentum dependence of energy is nonparabolic. We do
not believe however that these are crucial differences, and indeed the
small-$U$ limit mean-field solution \cite{kulik2} corresponds to an energy
gap $\Delta_0\simeq 12t\exp(-12t/U)$.
Figure \ref{cls},a shows minimal energy $E_N$ of cubic cluster as a
function of number of electrons $N$. Parity effect clearly
manifests itself in lower $E_N$ values for even $N$ and, when
plotting $\Delta_P$ against $\delta/\Delta_0$ (Fig.\ref{cls},b), displays a behavior
very similar to one received in the Richardson-Sherman model.
\ack
One of the authors (I.O.K.) acknowledges helpful
discussions  with the participants of International
Symposium on Mesoscopic Superconductivity (NTT,
March 2000, Atsugi, Japan),  especially with
A.F. Andreev, H. Takayanagi and R.I. Shekhter.
This work was partially supported by the Scientific
and Technical Research Council of Turkey (TUBITAK)
under grant No. TBAG 1736 and by the National
Research Council of Italy, under the Research and
Training Program for the Third Mediterranean
Countries.


\begin{thebibliography}{99}
\bibitem{anderson}
P.W. Anderson,
{\em J. Phys. Chem. Solids} {\bf 11} (1959) 28.
\bibitem{black}
C.T. Black, D.C. Ralph, and M. Tinkham,
{\em Phys. Rev. Lett.} {\bf 76} (1996) 688.
\bibitem{ralph}
D.C.Ralph, C.T.Black, J.M.Hengenrotter, J.G. Lu, and
M.Tinkham, in: Mesoscopic Electron Transport, p.447. Eds.
L.L.Sohn, L.P.Kouwenhoven and G.Sch{\"o}n. Kluwer NATO ASI Series,
vol.345, 1997.
\bibitem{matveev}
K.A. Matveev and A.I. Larkin,
{\em Phys. Rev. Lett.} {\bf 78} (1997) 3749.
\bibitem{braun}
F. Braun and J. von Delft,
{\em Phys. Rev. B} {\bf 59} (1999) 9527.
\bibitem{ring}
P. Ring and P. Shuck:
{\em The Nuclear Many-Body Problem}
(Springer-Verlag, New York, 1980).
\bibitem{kulik1}
I.O. Kulik and I.K. Yanson: {\em The Josephson Effect in
Superconductive Tunneling Structures} (Izdatel'stvo Nauka, Moscow,
1970; Keter Press, Jerusalem, 1972)
\bibitem{richardson}
R.W. Richardson and N. Sherman,
{\em Nuclear Physics} {\bf 52} (1964) 221.
\bibitem{delft}
J. von Delft and F. Braun,
{\em cond-mat/9911058}
\bibitem{boyaci}
H. Boyaci, Z. Gedik, and I.O. Kulik,
{\em submitted to Phys. Rev. B} and
{\em cond-mat/9909386}.
\bibitem{mastellone}
A. Mastellone, G. Falci, and
R. Fazio, {\em Phys. Rev. Lett.} {\bf 80} (1998) 4542.
\bibitem{berger}
S.D. Berger and B.I. Halperin,
{\em Phys. Rev. B} {\bf 58} (1998) 5213.
\bibitem{kulik2}
I.O. Kulik, {\em Physica} {\bf 126 B} (1984) 280.
\end{thebibliography}
\end{document}